\documentclass[twocolumn,prb,aps,bm,amssymb,amsmath,floats,showpacs]{revtex4}
\usepackage{epsf}
\begin{document}
\hsize\textwidth\columnwidth\hsize\csname @twocolumnfalse\endcsname    
\title{Crossover from tunneling to incoherent (bulk) transport in
a correlated nanostructure}

\author{ J.~K. Freericks }

\affiliation{
Department of Physics, Georgetown University, 
 Washington, D.C. 20057-0995, U.S.A.  }

\begin{abstract}
{
We calculate the junction resistance for a metal-barrier-metal device with
the barrier tuned to lie just on the insulating side of the metal-insulator
transition.  We find that the crossover from tunneling behavior in thin
barriers at low temperature to incoherent transport in thick barriers
at higher temperature is governed by a generalized Thouless energy.  The
crossover temperature can be estimated from the low temperature resistance
of the device and the bulk density of states of the barrier.
}
\end{abstract}


\maketitle

Many electronic devices employ quantum-mechanical tunneling in
determining their transport properties.  Examples include Josephson
junctions~\cite{josephson} and magnetic tunnel junctions~\cite{magnetic}.  
When designing a device
manufacturing process, or when optimizing the operational characteristics
of a device, it is important to have diagnostic tools that can determine 
if the transport is via tunneling or via defects in the barrier (such as
pinholes).  In superconductor-based devices, this is well understood, and was
described in detail by Rowell~\cite{rowell} 
in the 1970's.  However, the criteria relied
on testing the device in the superconducting state. Interest in this problem for
normal metals and for higher device operating temperatures has been driven
by recent activity in magnetic tunnel junctions~\cite{magnetic}.  A number of 
useful criteria for tunneling~\cite{schuller}
have emerged for these normal-metal-based devices: (i) the junction
resistance should increase with decreasing temperature; (ii) the fit of an
$IV$ characteristic to a Simmons model~\cite{simmons}
should have a barrier height that
does not decrease and a fitted thickness that does not increase as $T$
decreases; and (iii) the junction noise should not increase at finite
bias.  It has also been well established that the naive criterion for
tunneling, that the
resistance increases exponentially with the barrier thickness is insufficient,
since a rough interface plus pinholes will also yield this exponential
dependence~\cite{pinholes}.

In this contribution, we perform a theoretical analysis of tunneling through
a correlated barrier to investigate the crossover from a tunneling regime,
where transport is dominated by quantum processes that provide ``shorts'' 
across the barrier, to an incoherent bulk transport regime, where the
transport occurs via incoherent
thermal excitations of carriers in the barrier.  In the
latter case, one expects the junction resistance to scale linearly with the 
barrier thickness, with the slope proportional to the bulk resistivity of the
barrier (which has a strong temperature dependence in an insulator). As the 
barrier is made thinner (or the temperature is decreased), the direct 
quantum-mechanical coupling of the metallic leads through states localized
in the barrier begins to dominate the transport process, and the resistance
is reduced from that predicted by the incoherent transport mechanism to a
relatively temperature independent tunneling-based resistance.  Since the 
wavefunctions that connect the two metallic leads decay exponentially in the
barrier, the tunneling resistance depends exponentially on the barrier
thickness.  Most commercial devices operate in this tunneling regime because the
junction resistance is low enough to generate reasonable current values for 
low voltages and because the weak temperature dependence simplifies
variations of the device parameters with temperature.

In conventional tunneling devices, which use an insulator with a large
energy gap (like AlO$_x$), one cannot see the crossover to the bulk
transport regime, because it occurs at too high a temperature, or for too
resistive junctions to be of interest.  But there has been recent work in
examining barriers that are tuned to lie closer to the metal-insulator
transition~\cite{newman} 
(like Ta$_x$N), and thereby have much smaller ``energy gaps''.
Barriers of this type may be easier to work with because they can be made 
thicker and thereby be less susceptible to pinhole formation.  They also
can be advantageous for different applications.  As the energy gap of the 
barrier
material is made smaller (or equivalently, if the barrier potential height is
reduced), it becomes possible to observe and study the 
crossover from tunneling to bulk transport.

We consider a device constructed out of stacks of infinite two-dimensional
planes stacked in registry on top of each other.  This kind of inhomogeneous 
layered device can be used to describe a wide range of different 
multilayer-based structures.  We couple a bulk ballistic semi-infinite
metal lead to thirty self-consistent ballistic metal planes; then we stack
1 to 20 barrier planes and then top with another thirty self-consistent 
ballistic metal planes followed by another bulk ballistic semi-infinite metal
lead. The ballistic metal is described by a simple hopping Hamiltonian
with no interactions.  The barrier is described by a spin-one-half
Falicov-Kimball model~\cite{falicov_kimball}
with the same hopping parameters as the metal
plus strong scattering that yields correlations for the electron motion. 
The Hamiltonian is
\begin{equation}
{\mathcal{H}}=-t\sum_{\langle i,j\rangle\sigma}c^\dagger_{i\sigma}c_{j\sigma}
+\sum_{i\sigma}U_i^{FK}w_i(n_{i\sigma}-\frac{1}{2}),
\label{eq: ham}
\end{equation}
where $c^\dagger_{i\sigma}$ ($c_{i\sigma}$) creates (destroys) a conduction
electron at site $i$ with spin $\sigma$ and $t$ is the hopping parameter.
The hopping is on a simple cubic lattice 
constructed from the stacked two-dimensional planes; i.e., the hopping integral
is chosen to be the same within a plane and between two planes. $U_i^{FK}$ is 
the Falicov-Kimball interaction and $w_i$ is a classical variable, equal to zero
or one, which denotes the presence of a scatterer at site $i$.  Finally,
$n_{i\sigma}=c^\dagger_{i\sigma}c_{i\sigma}$ is the electron number operator.
The Falicov-Kimball interaction is nonzero only within the barrier, where
we set it equal to $6t$---large enough to create an insulator with a gap of
$0.4t$.  The average concentration of scatterers is $\langle w_i\rangle=1/2$
and we choose half filling for the electrons as well (with our choice of
interaction, this corresponds to a vanishing chemical potential). In order to be
quantitative, we pick the hopping parameter to satisfy $t=0.25$~eV, which
yields a bandwidth of 3~eV for the metallic leads and a gap of 100~meV for
the correlated insulator (much smaller than a conventional oxide 
insulator).  We solve for the Green's functions using an
inhomogeneous dynamical mean field theory calculation described 
elsewhere~\cite{potthoff_nolting,miller,jkf_scms,jkf_review}.
The resistance-area product for this device is calculated by a real-space 
version of Kubo's formula.  We take the lattice constant to be 0.3~nm.

\begin{figure}[htbf]
\epsfxsize=3.0in
\centerline{\epsffile{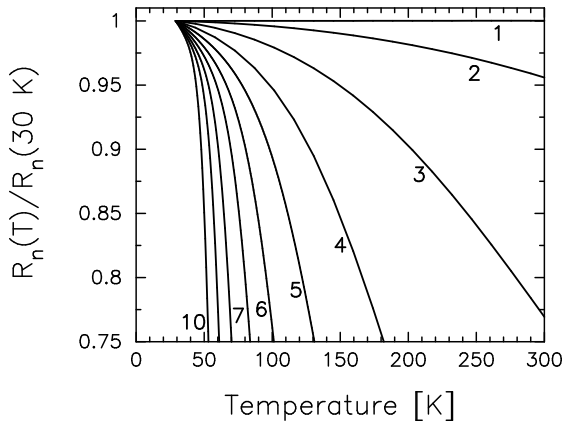}}
\caption{Ratio of the resistance of the junction at temperature $T$ to the
resistance at 30~K.  The different curves correspond to different thicknesses
of the barrier, which are labeled with an integer denoting the number of
atomic planes in the barrier.  As expected, the temperature dependence of the
resistance increases as the barrier is made thicker, because the barrier is
becoming more bulk-like.  However, in this regime, all of the transport is
still dominated by tunneling.
\label{fig: rn_t}
}
\end{figure}

In Fig.~\ref{fig: rn_t}, we plot the ratio of the junction resistance 
at temperature $T$ to the resistance at 30~K for junctions with a barrier
thickness ranging from 1 to 10 atomic planes. In all cases, the resistance
shows a weak temperature dependence with an insulator-like character.  This 
low-$T$ behavior is often used as a diagnostic to indicate that tunneling
is occuring in a junction~\cite{schuller,rudiger}, and that certainly is the 
case here.
Note how the temperature dependence increases as the thickness increases.
This is because the thicker the barrier is, the more it looks like
a bulk material, and an insulating barrier has strong (exponentially activated)
temperature dependence in the bulk.

\begin{figure}[htbf]
\epsfxsize=3.0in
\centerline{\epsffile{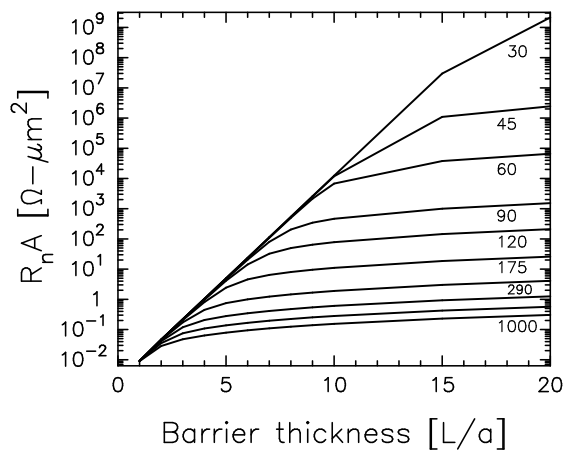}}
\caption{Resistance-area product as a function of the barrier thickness $L$
for a number of different temperatures (the labels on the curves are
in K).  Notice how the thin barriers have an
exponential dependence on thickness, which gives way to a linear dependence
as the junctions are made thick enough.  This crossover region moves to thinner
barriers as the temperature is increased.
\label{fig: rn_l}
}
\end{figure}

Since our junctions are defect free, with atomically smooth interfaces, we
can analyze the resistance at fixed temperature as a function of the
barrier thickness to look for exponential dependence in the tunneling
regime, with a crossover to linear dependence in the incoherent (bulk)
transport regime.  This is plotted in Fig.~\ref{fig: rn_l} for a number
of different temperatures, ranging from 30~K to 1000~K. Note how we see
a perfect exponential dependence on thickness for thin barriers, which then
gives way to a crossover to linear behavior as the junctions are made thicker
and the transport becomes incoherent and thermally activated.  Because
of the thermal activation, this crossover moves to thinner barriers as the
temperature is increased.  But it is interesting to note that there is no
simple relationship between the bulk gap (approximately 50~meV or 550~K when
measured from the $T=0$ chemical potential) and the location of the 
crossover thickness as a function of temperature.  Indeed, as $T$ is
increased, this crossover region is pushed to thinner and thinner
barriers.  This type of behavior has been seen in Josephson junctions
made from high temperature superconductors using 
molecular-beam-epitaxy~\cite{eckstein}.
When the barrier was increased from 1 to 3 to 5 to 7 atomic planes, the
junction resistance initially increased exponentially, and then started
to turn over to a more linear dependence on thickness.  However, because the
high-temperature superconductor is a d-wave superconductor, there is
strong temperature dependence to the junction resistance, even in the
tunneling regime, so direct comparison with results given here is impossible.
This behavior has also been seen in some magnetic tunnel 
junctions~\cite{covington} where an exponential increase as a function of
thickness gives way to an essentially constant dependence on thickness
for thicker Aluminum regions.  What is less known about this data is how much
of the Aluminum is oxidized in the manufacturing process.  Also, no 
temperature scans at fixed thickness were reported.

\begin{figure}[htbf]
\epsfxsize=3.0in
\centerline{\epsffile{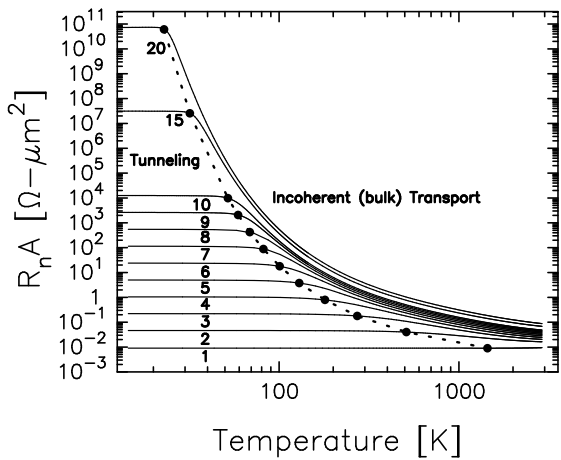}}
\caption{Resistance-area product as a function of temperature
for a number of different barrier thicknesses plotted on a log-log
plot.  Notice how the thin barriers have a weak dependence on temperature,
and a constant step-size increase in the logarithm of the resistance as
the thickness increases,
indicating tunneling behavior, and how there is a crossover to incoherent
transport as $T$ is increased.  The dashed line shows the boundary where
the generalized Thouless energy is equal to $k_BT$.  This marks the
approximate crossover from tunneling (for $E_{\rm{Th}}(T)\gg k_BT$) and
incoherent transport (for $E_{\rm{Th}}(T)\ll k_BT$).
\label{fig: thouless}
}
\end{figure}

In Fig.~\ref{fig: thouless}, we plot $R_n(T)A$ versus $T$ for a variety of
barrier thicknesses on a log-log plot.  This figure clearly shows the
tunneling regime, where the resistance-area product is approximately
constant, and it shows the incoherent regime, where the resistance-area product
has a strong temperature dependence.  The dashed line, that divides these two
regions is an approximate boundary that denotes the crossover region for 
the two different types of transport.  This crossover line is determined
by equating an energy scale extracted from the resistance with the temperature.
When this energy scale is larger than $k_BT$ we have tunneling, when it is lower
than $k_BT$ we have incoherent transport. The energy scale is a generalized
Thouless energy~\cite{thouless}, valid for a barrier that is described by an 
insulator that does not have either ballistic or diffusive transport. The 
generalized Thouless energy $E_{\rm{Th}}$ is the energy scale constructed 
from the resistance at temperature $T$ via the expression
\begin{equation}
E_{\rm{Th}}(T)=\frac{\hbar}{R_n(T)\frac{2e^2}{\hbar}\int d\omega
\left (-\frac{df(\omega)}{d\omega}\right ) \rho_{int}(\omega)L}
\label{eq: thouless}
\end{equation}
where $f(\omega)=1/[1+\exp(\omega/k_BT)]$ is the Fermi-Dirac distribution,
$\rho_{int}(\omega)$ is the bulk density of states in the insulator, and
$L$ is the barrier thickness.
This definition of $E_{\rm{Th}}$ agrees with the conventional notion of
$\hbar/t_{dwell}$, relating the Thouless energy to the dwell time in
the barrier, when the transport in the barrier is described by either
a ballistic metal
(where the Thouless energy varies like $C/L$) or a diffusive metal
(where the Thouless energy varies like $C/L^2$),
but it can now be generalized for an insulating barrier as well (where the
Thouless energy now picks up a substantial temperature dependence).

The notion of a Thouless energy can be employed as a diagnostic for tunneling
devices. Since $R_n(T)$ depends weakly on $T$ in the tunneling regime, 
one can measure $R_n$ at low $T$, and estimate the crossover temperature,
by computing a simple integral over the bulk insulator density of states.
Then one evaluates $E_{\rm{Th}}(T)$ employing Eq.~(\ref{eq: thouless}) using
the low-temperature value of the resistance.  The crossover temperature
is estimated by the point where $E_{\rm{Th}}(T)=k_BT$.
\textit{Note further that this crossover temperature is not proportional to the
gap of the bulk insulator, but rather is a complicated function of the
barrier thickness, and the strength of the correlations.}

In summary, we have determined an energy scale extracted from the resistance of 
a junction, that governs the crossover from tunneling to incoherent
transport. This energy scale approaches zero as the barrier thickness
becomes large, hence it could have applicability to any tunneling-based
device, but when we examine the common resistance-area products of actual
devices, it becomes clear that this concept will have the most applicability
to junctions with barriers tuned to lie close to the metal-insulator transition.
Since it is possible such devices will be used for devices of the future,
the concept of a generalized Thouless energy should become an important 
diagnostic tool in evaluating the quality of devices, and allow one to engineer
the thickness and operating temperature range to guarantee tunneling with
the chosen barrier.

{\it Acknowledgments:}
We acknowledge support from the National Science Foundation under grant
number DMR-0210717 and from the Office of Naval Research under grant number
N00014-99-1-0328.  High performance computer time was provided by the Arctic 
Region Supercomputer Center and the U. S. Army Engineering Research and 
Development Center. We also acknowledge useful discussions 
with  J. Eckstein, B. Jones, N. Newman, B.  Nikoli\'c, S. Parkin, J. Rowell, 
I. Schuller and S. Shafraniuk.

\end{document}